\documentclass[onecolumn,notitlepage,showpacs,amsmath,amssymb,prd,floatfix]{revtex4-1}
\usepackage{graphicx}
\usepackage{bm}
\usepackage{amsmath}
\usepackage{color}
\usepackage[dvipdfm,
 bookmarkstype=toc=true,
 linktocpage=true,
 bookmarks=true
 ]{hyperref}
\usepackage[varg]{txfonts}
\usepackage{natbib}
\definecolor{darkred}{cmyk}{0,1.00,1.00,0.3}

\newcommand{\hidden}[1]{}
\newcommand{\Ca}{\mathcal{C}}
\newcommand{\Ta}{\mathcal{T}}

\newcommand{\lin}{{\rm L}}
\newcommand{\non}{{\rm N}}

\newcommand{\mnras}{Mon.~Not.~Roy.~Astron.~Soc.}
\newcommand{\physrep}{Phys.~Rep.}

\newcommand{\apjl}{\apj~Lett.}

\def\ave#1{\left\langle #1 \right\rangle}

\newcommand{\simgt}{\lower.5ex\hbox{$\; \buildrel > \over \sim \;$}}
\newcommand{\simlt}{\lower.5ex\hbox{$\; \buildrel < \over \sim \;$}}

\newcommand{\tdelta}{\tilde{\delta}}
\newcommand{\tW}{\tilde{W}}

\newcommand{\bx}{{\bf x}}
\newcommand{\bk}{{\bf k}}
\newcommand{\btheta}{{\bm{\theta}}}
\newcommand{\bq}{{\bf q}}
\newcommand{\bl}{{\bf l}}
\newcommand{\tu}{\tilde{u}}
\newcommand{\wt}{\mathcal{W}}
\newcommand{\twt}{\tilde{\mathcal{W}}}
\newcommand{\tsigma}{\tilde{\Sigma}}

\begin{document}

\title[Power spectrum covariance]
{Power Spectrum {Super-}Sample Covariance}
\author{Masahiro Takada}
\affiliation{Kavli Institute for the Physics and Mathematics of the Universe
(Kavli IPMU, WPI),
The University of Tokyo, Chiba 277-8583, Japan}

\author{Wayne Hu}
\affiliation{
Kavli Institute for Cosmological Physics, Enrico Fermi Institute,
University of Chicago, Chicago, IL 60637\\
Department of Astronomy and Astrophysics, University of Chicago, Chicago, IL 60637}

\begin{abstract}
We provide a simple, unified approach to describing the impact of {super}-sample covariance, or beat coupling, on power
spectrum estimation in a finite-volume survey.
For a wide range of survey volumes, the sample
variance that arises from modes that are larger than the survey dominates the covariance of power
spectrum estimators for modes much smaller than the survey.   
The deeply
nonlinear version of 
this effect 
is
known as halo sample variance.
We show that  all variants are unified by the matter trispectrum of squeezed configurations and that
such configurations obey a consistency relation which relates them to the response of the
power spectrum to a change in the background density. Our method also applies to statistics that are based
on radial projections of the density field such as weak lensing shear.  While we use the halo model for
an analytic description to expose the nature of the effect, 
 the consistency description enables an accurate calibration of the
full effect directly from simulations.  It also suggests that {super}-sample covariance may be
viewed as an additional interesting signal rather than excess noise.
\end{abstract}

\maketitle

\section{Introduction}

Cosmic acceleration is perhaps the most tantalizing
problem in modern cosmology. To unlock its mysteries, a number of ambitious wide-field optical and infrared
galaxy surveys have been proposed. These range from ground-based imaging
and spectroscopic surveys such as the Subaru Hyper Suprime-Cam 
(HSC)
Survey \footnote{\url{http://www.naoj.org/Projects/HSC/index.html}}, 
the Dark Energy Survey (DES) \footnote{\url{http://www.darkenergysurvey.org}}, 
the Kilo-Degrees Survey
(KIDS) \footnote{\url{http://www.astro-wise.org/projects/KIDS/}}, the
LSST \footnote{\url{http://www.lsst.org/lsst/}}, the Baryon Oscillation
Spectrograph Survey
(BOSS) \footnote{\url{http://cosmology.lbl.gov/BOSS/}}, 
the Extended BOSS (eBOSS) 
\footnote{\url{http://www.sdss3.org/future/eboss.php}}, 
the BigBOSS \footnote{\url{http://bigboss.lbl.gov}}, and the Subaru
Prime Focus Spectrograph (PFS)
Survey \footnote{\url{http://sumire.ipmu.jp/pfs/intro.html}}\cite[see also][]{Ellisetal:12} to
space-based optical and near-infrared missions such as 
the Euclid project
\footnote{\url{http://sci.esa.int/science-e/www/area/index.cfm?fareaid=102}}
and the WFIRST project \footnote{\url{http://wfirst.gsfc.nasa.gov}}.  
Each of these surveys approaches the nature of cosmic acceleration using
multiple large-scale structure probes: e.g.~weak gravitational lensing, baryon
acoustic oscillations, and clustering statistics of large-scale structure
tracers such as galaxies and clusters, including redshift-space distortion effects
\citep[see Ref.][for a recent review]{Weinbergetal:12}.

To attain the full potential of such surveys, it is important to
understand the statistical properties of large-scale structure probes and the
matter density field that underlies them.
Its two-point
correlation function or the Fourier-transformed counterpart, the power
spectrum, is the simplest and most commonly used statistical quantity to
extract cosmological information from the large-scale structure probes. 
The statistical precision of  power 
{spectrum}
measurements is determined by their
covariance matrix that itself contains two contributions; the
measurement noise  and  sample
variances caused by 
{an}
incomplete sampling of the fluctuations due to a
finite-volume survey.

Even though the initial density field is nearly Gaussian, the sample variance
of large-scale structure probes gets substantial non-Gaussian contributions from 
the nonlinear evolution of 
{large-scale structure}
\cite{Scoccimarroetal:99,HuWhite:01,CoorayHu:01}. 
Most of the useful cosmological information from such
probes lies in the weakly or deeply nonlinear regime in the matter
distribution. In the nonlinear regime, the different Fourier modes are
no longer independent but rather are correlated with each other. 
Hence, 
off-diagonal entries in the power spectrum covariance matrix appear which are 
described by the connected 4-point correlation function of the matter distribution,
the matter trispectrum.
 To realize the statistical power carried
by power spectrum, it is important to accurately model the matter
trispectrum in the nonlinear regime.
The standard methods to study the power
spectrum sample variance are based on simulations of the
$\Lambda$-dominated cold dark matter ($\Lambda$CDM) model
\cite{Scoccimarroetal:99,HuWhite:01,Hamiltonetal:06,Sefusattietal:06,Sembolonietal:07,Satoetal:09,Takahashietal:09,Yuetal:12,Kayoetal:12,Pen:12,Maneraetal:13}
which can be used to test and calibrate semi-analytic descriptions such as the 
halo model approach that can encompass a larger range of model parameters
\cite{CoorayHu:01,TakadaBridle:07,Neyrincketal:07,TakadaJain:09,Kayoetal:12}.

One further 
 complication arises. Nonlinear evolution couples short-wavelength modes relevant for the
power spectrum measurement with the very long-wavelength modes outside
 a survey volume, the so-called super-survey
modes. The super-survey modes are tricky to consider, because their effect
vanishes for power spectrum measurements of simulations with periodic boundary
conditions that have no contribution of modes outside the simulation box.
Super-sample covariance on the power spectrum
was originally pointed out in Ref.~\cite{Hamiltonetal:06} and called beat coupling
 \citep[see also][which studied the super-sample variance
 effect on the
 number counts of halos]{HuKravtsov:03}, and then was followed by many
 studies 
\cite{Sefusattietal:06,TakadaBridle:07,TakadaJain:09,Satoetal:09,Takahashietal:09,dePutter:2011ah,Kayoetal:12}.
Later the effect of super-survey modes for power spectra in the deeply nonlinear regime was studied using halo bias theory \cite{Moetal:97} in the halo
model approach and called halo sample variance 
\cite{TakadaBridle:07,Satoetal:09,Kayoetal:12} 
\citep[see also
][]{HuKravtsov:03}. 
These
studies have shown that super-sample variance 
dominates the non-Gaussian errors in the weakly or deeply nonlinear regime.
Super-sample variance contributes to
 a saturation in the information content carried by the power
spectrum amplitude
\cite{Hamiltonetal:06,Satoetal:09,Zhangetal:11,Yuetal:12,Kayoetal:12}
\citep[see also][for tests on real data]{LeePen:08}.
The saturated information content has triggered a further discussion on
how to recover the information content that was originally contained in
the initial Gaussian field, by using the nonlinear transformation method
\citep[e.g.][]{Neyrincketal:09,Seoetal:11} or including the higher-order
moment information \cite{TakadaJain:04,Kayoetal:12}.

However, it is still unclear how all of these super-sample variance effects are  
described
by the matter trispectrum and how they can best be quantified and
treated in parameter estimation.
In this paper, we provide a systematic study of common origin of all such effects,
unify their description, and show how they can be directly quantified.
%
%
%

The structure of this paper is as follows. In \S~\ref{sec:formula}, we
develop a simple, unified approach to describe the super-sample covariance of
power spectrum estimation in a finite-volume survey and show its relation to the response of the
power spectrum to a change in the background density. In
\S~\ref{sec:halo}, we use the halo model to compute the power spectrum
covariance for a $\Lambda$CDM model and verify that we can recover the beat-coupling
and halo sample variance results from our description. \S~\ref{sec:discussion} is devoted
to discussion.  We provide an Appendix that extends the formalism to projected density field
statistics such as weak lensing shear.

\section{Power spectrum covariance in a survey window}
\label{sec:formula}

In \S \ref{sec:PSestimator} we review the construction of power spectrum
estimators in a  {finite-volume survey}
and discuss their relation to the underlying true power spectrum.   We relate the covariance
of these estimators to the matter trispectrum in \S \ref{sec:PScovariance} and identify where the
additional sample variance from modes larger than the survey arises.    In \S \ref{sec:consistency}, we
present a trispectrum consistency relation that straightforwardly relates this effect to the response
of the power spectrum to a change in the background density and outline how it can be easily calibrated in
simulations.  As such, it may also be considered as signal rather than noise, carried by a
single additional parameter, the variable background density.

\subsection{Power spectrum estimator}
\label{sec:PSestimator}

Assume
we measure the underlying density fluctuation field $\delta(\bx)$ through a survey window function
 $W(\bx)$
which is 1 in the measured region and 0 in the unmeasured region. 
Note that by construction we implicitly assume that the observable is the density fluctuation itself or
equivalently the true mean density is known.   This is appropriate for
statistics related to weak lensing that probes the differential
gravitational field, i.e.\ the tidal field.  If the
density fluctuation field is measured relative to the mean of the survey region there is a
correction factor that we will describe in \S \ref{sec:consistency} (see
also 
Ref.~\cite{dePutter:2011ah}).

Regardless of the complexity of the survey geometry, we can always define
the  observed field as
\begin{equation}
\delta_W(\bx) = \delta(\bx) W(\bx),
\end{equation}
whose Fourier transform is a convolution
\begin{equation}
\tdelta_W(\bk) = \int\!\! \frac{d^3 \bq}{(2\pi)^3} \tilde{W}(\bq)
 \tdelta(\bk-\bq), 
\end{equation}
where $\tilde{W}(\bk)$ is the Fourier transform of the survey window function. 
We have here employed the
continuous  limit of discrete Fourier transforms under the approximation
that the total volume {for the Fourier transform}
is much greater than the survey region 
(see
Ref.~\cite{TakadaBridle:07,Kayoetal:12} for a
pedagogical derivation of power spectrum estimator and the covariance
 based on the discrete Fourier decomposition).

Let us next define an estimator of the power spectrum as
\begin{equation}
\hat P(k_i) \equiv \frac{1}{V_W} \int_{|\bk|\in k_i}\!\! \frac{d^3
 \bk}{V_{k_i}} 
\tilde\delta_W(\bk)\tilde\delta_W(-\bk),
\label{eq:ps_est}
\end{equation}
where the integral is over a shell in $k$-space of width $\Delta k$ and
 volume 
$V_{k_i}\simeq 4\pi k_i^2 \Delta k$ for $\Delta k /k_i \ll 1$.
 Here the effective survey volume is defined as
\begin{equation}
 V_W\equiv \int\!d^3\bx~W(\bx).
\end{equation}
Given the definition of the power spectrum
\begin{equation}
\langle \tilde \delta(\bk ) \tilde \delta(\bk') \rangle = (2\pi)^3 \delta_D^3(\bk+\bk') P(k),
\end{equation}
the ensemble average of its estimator is 
\begin{eqnarray}
\langle \hat P(k_i) \rangle &=& \frac{1}{V_W}   \int_{|\bk|\in 
k_i}\!\! \frac{d^3 \bk}{V_{k_i}}  \int\!\! \frac{d^3 \bq}{(2\pi)^3}  
\left|\tilde{W}(\bq)\right|^2 P(\bk-\bq).
\end{eqnarray}
Thus, the observed power spectrum is given as a convolution of the
underlying power spectrum with the window function combining the density
modes separated by less than the Fourier width of the window.  
In the general case,
one would deconvolve the window in constructing an unbiased estimator
\cite{Tegmarketal:04,Hikageetal:11}.

In this paper, we are interested in the effect of the global survey
geometry on the power spectrum covariance, not in the effect of masked
regions at small spatial scales. In this case, the window function has a
width of $\sim 1/L$ in Fourier space, where $L \sim V_W^{1/3}$ or more generally its smallest dimension. When we focus on wavenumber modes satisfying
$k\gg 1/L$, we  find that the power spectrum estimator of Eq.~(\ref{eq:ps_est}) is unbiased:
\begin{eqnarray}
\langle \hat P(k_i) \rangle &\simeq& \frac{1}{V_W}\int_{|\bk|\in 
k_i}\!\! \frac{d^3 \bk}{V_{k_i}} P(k)  \int\!\! \frac{d^3 \bq'}{(2\pi)^3}  
\left|\tilde{W}(\bq')\right|^2
\simeq P(k_i)\frac{1}{V_W}\int\!\! \frac{d^3 \bq'}{(2\pi)^3}  
\left|\tilde{W}(\bq')\right|^2=P(k_i).
\label{eq:Pk}
\end{eqnarray}
Here we have used that $P(|\bk-\bq|)\simeq P(k)$ over the integration range
of $d^3\bq$ which the window function supports and also assumed that $P(k)$ is
not a rapidly varying function within the 
$k$-bin.  In the third
equality on the r.h.s., we have used the general identity for the window
{function}: 
\begin{equation}
V_{W}=\int\!d^3\bx~W^n(\bx)=
\int\!\left[\prod_{a=1}^n \frac{d^3\bq_a}{(2\pi)^3}
\tW(\bq_a)\right](2\pi)^3\delta^3_D(\bq_{1\ldots n}),
\label{eq:generalVW}
\end{equation}
where $\bq_{1 \ldots n}= \bq_1 +\ldots \bq_n$ here and below.
For $n=2$,
 $V_W=\int  |\tW(\bq)|^2 d^3\bq /(2\pi)^3$.

\subsection{Power spectrum covariance}
\label{sec:PScovariance}

Now consider the power spectrum covariance, which can be defined in
terms of the power spectrum estimator as
\begin{equation}
C_{ij}\equiv {\rm Cov}[P(k_i),P(k_j)]=
\ave{\hat{P}(k_i)\hat{P}(k_j)}-
\ave{\hat{P}(k_i)}\ave{\hat{P}(k_j)}.
\end{equation}
Here we consider sample covariance only. 
A real measurement will have measurement
noise covariance, but we do not consider its effect in this paper.
In the same $k\gg 1/L$ limit, the covariance becomes
\begin{equation} 
C_{ij} \simeq \frac{1}{V_W} \left[ \frac{(2\pi)^3}{V_{k_i}} 2 P(k_i)^2
			     \delta^K_{ij} + \bar{T}^W(k_i,k_j) \right],
\label{eq:pscov1}
\end{equation}
where $\delta^K_{ij}$ is the Kronecker delta function; $\delta_{ij}^K=1$
if $k_i=k_j$ to within the bin width, otherwise $\delta_{ij}^K=0$. The
second term, proportional to $\bar{T}^W(k_i,k_j)$, is the non-Gaussian
contribution arising from the connected 4 point function or trispectrum
\citep[see also][for
the similar derivation]{Hamiltonetal:06},
\begin{equation}
\langle \tilde \delta(\bk_1 ) \tilde \delta(\bk_2 )\tilde \delta(\bk_3 )\tilde \delta(\bk_4 )\rangle_c =
(2\pi)^3 \delta_D^3(\bk_{1234}) T(\bk_1,\bk_2,\bk_3,\bk_4),
\end{equation}
convolved with the survey
window function:
\begin{equation}
 \bar{T}^W(k_i,k_j)=\frac{1}{V_{W}}
\int_{|\bk|\in k_i}\!\!\frac{d^3\bk}{V_{k_i}}
\int_{|\bk|'\in k_j}\!\!\frac{d^3\bk'}{V_{k_j}}
\int\!\left[\prod_{a=1}^4\frac{d^3\bq_a}{(2\pi)^3}
\tW(\bq_a)\right](2\pi)^3\delta^3_D(\bq_{1234})
T(\bk+\bq_1,-\bk+\bq_2,\bk'+\bq_3,-\bk'+\bq_4).
\label{eq:pscov}
\end{equation}
The convolution with the window function means that 
 different 4-point configurations separated by less than the Fourier
width of the window function and involving  contributions from super-survey modes
contribute in principle. 
{In deriving
Eq.~(\ref{eq:pscov1}),  we have again used the general window function identity
Eq.~(\ref{eq:generalVW}) with $n=4$ to eliminate one factor of $1/V_W$
in the 
power spectrum term under the same
approximation as in Eq.~(\ref{eq:Pk}).} 
{If}
the same slowly-varying approximation were true of the trispectrum, 
it can be taken out
of the integral and we would obtain the standard result
\cite{Scoccimarroetal:99} 
%
\begin{equation}
C_{ij} \approx
\frac{1}{V_W}  \frac{(2\pi)^3}{V_{k_i}} 2 P^2(k_i)
			     \delta^K_{ij}  + C_{ij}^{T0},
\label{eq:pscov_org}
\end{equation}
where
\begin{equation}
C_{ij}^{T0} =
\frac{1}{V_W}
\int_{|\bk|\in k_i}\!\frac{d^3\bk}{V_{k_i}}
\int_{|\bk'|\in k_j}\!\frac{d^3\bk'}{V_{k_j}}
T(\bk,-\bk,\bk',-\bk') ,
\label{eq:pscov_T0}
\end{equation}
so that the whole covariance scales as $1/V_W$.   Furthermore the first term can be understood from
Gaussian statistics: 
 the number of independent $k$-modes in the shell is given as
\begin{equation}
 N_{\rm mode}(k_i)=\frac{V_{k_i} V_W}{2(2\pi)^3}\simeq 
 \frac{2\pi k_i^2\Delta kV_W}{(2\pi)^3}
\end{equation}
such that 
\begin{equation}
 C^{\rm G}_{ij}=\frac{1}{N_{\rm mode}(k_i)}P^2(k_i)\delta^K_{ij}.
\end{equation}
%
The factor $2$ in the denominator of $N_{\rm mode}$ arises from
the reality condition of the density field,
i.e.\ $\tdelta_W^\ast(\bk)=\tdelta_W(-\bk)$. The  difference in bin width $\Delta k$ scaling between the
Gaussian and non-Gaussian terms means that the latter becomes more prominent in the variance  when averaging over wide bins or equivalently when considering
the covariance of narrow bins.

For the non-Gaussian term,
there are additional effects if the trispectrum at $k \gg 1/L$ has structure on the scale of the survey
$\Delta k \lesssim 1/L$.     We next consider the general origin and description of such a term.

\subsection{Super-sample covariance 
and trispectrum consistency}
\label{sec:consistency}

The trispectrum term that governs the additional effects are so-called squeezed quadrilaterals where
two pairs of sides are nearly equal and opposite.
By analogy with similar effects for primordial non-Gaussianity \cite{Maldacena:03},
these configurations should be determined by the response of the power spectrum to
a rescaling of the background.   We call this separate-universe ansatz the
{\em trispectrum 
consistency relation}.

To see this
in Eq.~(\ref{eq:pscov}),
we can make the change of variables $\bk+\bq_1\leftrightarrow \bk$ and 
$\bq_1+\bq_2\leftrightarrow \bq_{12}$ under the delta function
condition $\bq_{1234}={\bf 0}$ and the approximation that $q_{12} \ll k, k'$.   The 
term of interest therefore is 
\begin{equation}
\lim_{q_{12} \rightarrow 0} T(\bk,-\bk+\bq_{12},\bk',-\bk'-\bq_{12}).
\end{equation}
In this limit, the 4 point configuration describes the connection between $P(k)$ and $P(k')$ 
through a shared infinite wavelength mode $\bq_{12}$.  This mode acts like a background density
or constant mode to the short wavelengths $\bk$ and $\bk'$.   It follows therefore that 
the squeezed trispectrum can be characterized by the response of $P(k)$
to a fluctuation in the
background density $\delta_b$
through 
\begin{equation}
\bar T(\bk,-\bk+\bq_{12},\bk',-\bk'-\bq_{12})
\approx  {T}(\bk,-\bk,\bk',-\bk')+
\frac{\partial P(k)}{\partial \delta_b} \frac{\partial P(k')}{\partial \delta_b} P^{\lin}(q_{12}) .
\label{eq:consistency}
\end{equation}
The overbar here refers to an angle average over the direction of 
$\bq_{12}$ since any directional
dependence cannot be quantified by a purely constant mode.
Here $P^\lin$ is the linear power spectrum and is designated as such to remind the reader
that for this relation to be applicable  $\delta_b$ must be  a mode in
the linear regime. 
On the other hand, there is no such restriction on the $k$-modes of $P(k)$. 


In terms of the power spectrum covariance, this relation has the direct interpretation that the
measured power in the $k$ and $k'$ bins are correlated by the underlying background fluctuation
$\delta_b$ that they share.   
Trispectrum consistency then implies that the covariance is
\begin{eqnarray}
 C_{ij} = C_{ij}^{\rm G} + C_{ij}^{T0} 
 %
%
+(\sigma_W^L)^2  \frac{\partial   P(k_i)}{\partial  \delta_b} \frac{\partial  P(k_j)}{\partial  \delta_b} ,
\label{eq:covcon}
\end{eqnarray}
where we have introduced the variance of the background density
field $\delta_b$ in the survey window, defined as
\begin{equation}
(\sigma_W^L)^2\equiv \frac{1}{V_W^2}
\int\!\!\frac{d^3\bq}{(2\pi)^3}
|\tW(\bq)|^2P^{\lin}(q).
\label{eq:sigmw}
\end{equation}
In reducing Eq.~(\ref{eq:pscov}) with the consistency trispectrum 
Eq.~(\ref{eq:consistency}) to Eq.~(\ref{eq:covcon}), 
we used the 
following identity for the window function from the convolution theorem:
\begin{equation}
\int\!\frac{d^3\bq_1}{(2\pi)^3}\tW(\bq_1)\tW(\bq-\bq_1)=
\int\!d^3\bx W(\bx)^2e^{i\bx\cdot\bq}
=\int\!d^3\bx W(\bx)e^{i\bx\cdot\bq}=\tW(\bq).
\label{eq:WWk}
\end{equation}
We call the $\delta_b$ type covariance term the super-sample covariance of the power spectrum estimator.
Note that because the background correlates changes in $\hat P(k)$ for all $k$, it can appear as
the dominant contribution to the (co)variance when the measurement involves a 
large number of $k$-modes.   It scales with the volume of the survey only through $(\sigma_W^L)^2$
whereas the other terms scale like white noise $V_W^{-1}$.   Until the survey becomes much larger
than the turnover of matter power spectrum on the horizon scale of matter-radiation equality, its relative
contribution remains important \cite{HuKravtsov:03,Kayoetal:12} 
(see also \S \ref{sec:LCDM}).

As in the case of super-survey mode effects for number counts \cite{Lima:2004wn}, their
impact on the power spectrum should perhaps not be considered as excess noise but rather a new signal: the fractional power spectrum response is a template that changes the measured power spectrum coherently across bins as
\begin{equation}
P(k) \rightarrow P(k) \left( 1 + \frac{\partial \ln P(k)}{\partial \delta_b} \delta_b \right) .
\end{equation}
The uncertainty is carried by a single unknown parameter $\delta_b$ drawn from
a Gaussian of variance $(\sigma_W^L)^2$.  Use of the covariance formalism in data analysis
would premarginalize over this parameter,
 but it may alternately be a fitting parameter for the power spectrum that has a prior given by $\sigma_W^L$.    This point of view may also be more appropriate for surveys that are not volume limited, i.e.\ where the window is not strictly 0 or 1.   Loss of
information on other cosmological parameters of interest that also
change the power spectrum only comes through degeneracies with this
single mode \citep[see also][for the similar discussion]{TakadaBridle:07,TakadaJain:09}. 


It is also easy in this language to account for the alternate  definition of the
density fluctuation field as relative to the mean of the survey region.   In this case
the observed power would be rescaled as $ P_W(k) =  P(k)/(1+\delta_b)^2$ and
the trispectrum consistency would take the same form as 
{Eq.~(\ref{eq:consistency}) and
(\ref{eq:covcon})}
but with the response to the background mode altered to be
\begin{equation}
\frac{\partial P(k)}{\partial \delta_b} \rightarrow
\frac{\partial  P_W(k)}{\partial \delta_b} \approx
 \frac{\partial  P(k)}{\partial \delta_b}- 2 P(k),
\end{equation}
which generalizes the treatment in Ref.~\cite{dePutter:2011ah}.  Given that the
fractional response function $\partial \ln P(k)/\partial \delta_b$ is typically positive and
order unity, the response in $P_W(k)$ can be significantly reduced compared with $P(k)$. Note again that if the observable of interest does not depend on the
mean density, which is the case for weak lensing shear that probes the
differential gravitational field (tidal field),  
 $P(k)$ rather than $P_W(k)$ is the relevant quantity. 

The fractional response function $\partial \ln P(k)/\partial \delta_b$ is a quantity that
can be directly calculated in simulations  since introducing $\delta_b$ is
equivalent to simulating an FRW universe with different background densities
 \cite{Sirko:05,Gnedin:2011kj,Baldauf:2011bh}.   For example, in principle for its evaluation at $z=0$ only two
 simulations would be required to calibrate all such effects  deep into the nonlinear regime.
 Nonetheless, we find it
illustrative to highlight the main features of power spectrum
super-sample covariance, as well as derive 
{Eq.~(\ref{eq:covcon})}
explicitly, analytically through the halo model. This will also allow us to make contact with the existing literature on this
effect,  ``beat coupling'' (BC) in second order perturbation theory
and
``halo sample variance'' (HSV) in the deeply nonlinear regime.
There are also terms related converting $P(k)$ to 
observables related to changes in the distance-redshift relation if the background mode encompasses the entire
volume out to the observer.  These in principle can affect the interpretation of baryon acoustic oscillation features (see \cite{SherwinZaldarriaga:12} for related sub-survey effects). 
We leave a more precise calibration and observational considerations to future work.

To be comprehensive, in  the Appendix we also develop the analogous
formulae to describe the power spectrum covariance for a two-dimensional,
density field that is obtained by projecting the three-dimensional
field, weighted with a selection function, along the line-of-sight.
{The formulation developed in this paper can be straightforwardly
extended to covariance theory for higher-order correlation functions
\citep[see][for the attempt to model the BC effect on the lensing
bispectrum covariance]{Kayoetal:12},
and the {super}-sample variance effects should be
similarly described by the response of the higher-order correlation to a
change in the background density.}

\section{Halo model approach}
\label{sec:halo}

We have seen in the previous section that the power spectrum estimator covariance in the survey window is generally described
by the matter trispectrum and the specific effects of super-sample covariance  by its squeezed
configurations.   The matter trispectrum itself can be approximated in the halo model by considering
correlations induced between dark matter halos via perturbation theory and within halos via
the universal density profile \cite{CoorayHu:01,CooraySheth:02}.   
Although the halo model is an
empirical model to describe the nonlinear clustering, it gives a fairly
accurate prediction compared to the simulations, e.g.\ to within a
10-20\% accuracy in the power spectrum amplitude at scales of interest
\cite{Kayoetal:12}. 
Evaluating the covariance using the halo model can therefore
illustrate the general consistency construction that the additional super-sample covariance terms
are equivalent to the response of the power spectrum to a background density mode.  We
develop the halo model formalism in \S \ref{sec:HMformalism} and illustrate it in the $\Lambda$CDM
context in \S \ref{sec:LCDM}.

\subsection{Formalism}
\label{sec:HMformalism}

In the halo model
 \cite{PeacockSmith:00,Seljak:00,MaFry:00,CooraySheth:02}, 
 the power spectrum itself is described as
\begin{equation}
P(k) = I_2^0(k,k) + [I_1^1(k)]^2 P^{\lin}(k),
\end{equation}
where the first term involves two points correlated by being in the
 same halo and the second two points in separate halos that are themselves correlated
 by the linear power spectrum.    We use the general notation \cite{CoorayHu:01}
 \begin{equation}
 I^\beta_\mu(k_1,k_2,\dots,k_\mu)\equiv 
\int\!\!dM\frac{dn}{dM}\left(\frac{M}{\bar{\rho}_m}\right)^\mu
b_\beta \tu_M(k_1)\tu_M(k_2)\cdots\tu_M(k_\mu),
\end{equation}
where $M$ is the halo mass, $dn/dM$ is the halo mass function, $b_0=1$,  $b_1=b(M)$ is the halo bias, and $\tilde u_M(k)$ is the Fourier transform of the halo density profile
normalized so that $\tilde u_M(0)=1$.  We have here assumed linear halo bias in that
$b_\beta=0$ for $\beta\ge 2$ \citep[see][for a possible extension of the halo model to
including the nonlinear halo bias]{CoorayHu:01}. This approximation does not
affect the main results of this paper. 
Note that for a halo bias that satisfies
the peak-background consistency relation
\begin{equation}
\int\!\!dM\frac{dn}{dM}\left(\frac{M}{\bar{\rho}_m}\right)
b(M) = 1,
\end{equation}
$I_1^1(0)=1$, the 2 halo term as $k \rightarrow 0$ is simply the linear power spectrum. 
Furthermore the same peak-background consistency of the bias
\cite{Moetal:97} says that
\begin{equation}
\frac{ \partial I_2^0}{\partial \delta_b} = I_2^1,
\label{eq:Iconsistency}
\end{equation}
which will be useful in relating the power spectrum covariance to its response to a background
mode $\delta_b$ (see \S \ref{sec:consistency}).

Likewise, the halo model approach tells us that the matter trispectrum arises from contributions involving one to four halos:
\begin{equation}
T=T^{1h}+\left(T^{2h}_{22}+T^{2h}_{13}\right)+T^{3h}+T^{4h}, 
\end{equation}
where $T^{1h}, \cdots, T^{4h}$ denote the 1-, 2-, 3- and 4-halo terms
that arise from correlations between four points that reside in the 1
halo (the same halo) and from 2 to 4 different halos,
respectively. Using the notations defined in Ref.~\cite{CoorayHu:01}, the
different halo terms are given as
\begin{eqnarray}
 T^{1h}(\bk_1,\bk_2,\bk_3,\bk_4)&=&
I^0_4(k_1,k_2,k_3,k_4),\nonumber\\
T^{2h}_{22}(\bk_1,\bk_2,\bk_3,\bk_4)&=&
P^{\lin}(k_{12})I^1_{2}(k_1,k_2)I^1_{2}(k_3,k_4)+
\mbox{2 perm.},\nonumber\\
T^{2h}_{13}(\bk_1,\bk_2,\bk_3,\bk_4)&=&
P^{\lin}(k_1)I_1^1(k_1)I^1_{3}(k_2,k_3,k_4)+\mbox{3 perm.},\nonumber\\
T^{3h}(\bk_1,\bk_2,\bk_3,\bk_4)&=&
B^{\rm PT}(\bk_1,\bk_2,\bk_{34})
I^1_1(k_1)I^1_1(k_2)I^1_2(k_3,k_4)
+\mbox{5 perm.},\nonumber\\
T^{4h}(\bk_1,\bk_2,\bk_3,\bk_4)&=&T^{\rm PT}(\bk_1,\bk_2,\bk_3,\bk_4)
I^1_1(k_1)I^1_1(k_2)I^1_1(k_3)I^1_1(k_4).
\end{eqnarray}
The 2-halo term has two contributions, $T^{2h}_{13}$ and $T^{2h}_{22}$ where one or two  point(s) among the four points are in the first halo,
with the remaining points in the second halo.
Here
$B^{\rm PT}$ and $T^{\rm
PT}$ are the matter bispectrum and trispectrum
given based on perturbation theory \cite{Bernardeauetal:02}
\begin{eqnarray}
B^{\rm PT}(\bk_1,\bk_2,\bk_3)&=&2F_2(\bk_1,\bk_2)P^{\lin}(k_1)P^{\lin}(k_2)
+\mbox{2 perm.},\nonumber\\
T^{\rm PT}(\bk_1,\bk_2,\bk_3,\bk_4)&=&
4\left[F_2(\bk_{13},-\bk_1)F_2(\bk_{13},\bk_2)
P^{\lin}(k_{13})P^{\lin}(k_1)P^{\lin}(k_2)+11 {\rm perm.}
\right]\nonumber\\
&&+6\left[F_3(\bk_1,\bk_2,\bk_3)
P^{\lin}(k_1)P^{\lin}(k_2)P^{\lin}(k_3)+\mbox{3 perm.}
\right],
\end{eqnarray}
where
\begin{align}
F_2(\bk_1,\bk_2) \equiv &\frac{5}{7}+\frac{1}{2}
\left(\frac{1}{k_1^2}+\frac{1}{k_2^2}\right)(\bk_1\cdot\bk_2)
+\frac{2}{7}\frac{(\bk_1\cdot\bk_2)^2}{k_1^2k_2^2},\nonumber\\
F_3(\bk_1,\bk_2,\bk_3) \equiv &
 \frac{7}{18}\frac{\bk_{12}\cdot\bk_1}{k_1^2}
 \left[F_2(\bk_2,\bk_3)+G_2(\bk_1,\bk_2)\right]
+\frac{1}{18}\frac{k_{12}^2(\bk_1\cdot\bk_2)}{k_1^2k_2^2}
\left[G_2(\bk_2,\bk_3)+G_2(\bk_1,\bk_2)\right], \nonumber\\
G_2(\bk_1,\bk_2)\equiv & \frac{3}{7}+\frac{1}{2}
\left(\frac{1}{k_1^2}+\frac{1}{k_2^2}\right)
(\bk_1\cdot\bk_2)+
\frac{4}{7}\frac{(\bk_1\cdot\bk_2)^2}{k_1^2k_2^2}
\end{align}
%
Strictly speaking, the mode-coupling kernels, $F_n$ and $G_n$, are
exact only for an Einstein de-Sitter model $(\Omega_{\rm m0}=1)$ and
have
a very weak dependence on the density parameters for models with
$\Omega_{\rm m0}\ne 1$
\citep{Bernardeauetal:02}. 
Here we employ the Einstein de-Sitter approximation for simplicity. This impacts how
we make the comparison to the response of the power spectrum to a background mode.

If $k, k'\gg q_i$  each halo term of the trispectrum can be
approximated as
\begin{eqnarray}
 T^{1h}(\bk,-\bk+\bq_{12},\bk',-\bk'-\bq_{12})
&\simeq & 
T^{1h}(k,k,k',k'), \nonumber\\
 T^{2h}_{22}(\bk,-\bk+\bq_{12},\bk',-\bk'-\bq_{12})&\simeq & 
T^{2h}_{22}(\bk,-\bk,\bk',-\bk')
+P^{\lin}(\bq_{12})I^1_2(k,k)I^1_2(k',k'),
\nonumber\\
 T^{2h}_{13}(\bk,-\bk+\bq_{12},\bk',-\bk'-\bq_{12})&\simeq & 
 T^{2h}_{13}(\bk,-\bk,\bk',-\bk'),
\nonumber\\
 T^{3h}(\bk,-\bk+\bq_{12},\bk',-\bk'-\bq_{12})
&\simeq& 
 T^{3h}(\bk,-\bk,\bk',-\bk')
\nonumber\\
&&
\hspace{-10em}+
4I^{1}_2(k,k)[I^1_1(k')]^2P^{\lin}(q_{12})P^{\lin}(k')F_2(\bq_{12},\bk')
+4I^{1}_2(k',k')[I^1_1(k)]^2
P^{\lin}(q_{12})P^{\lin}(k)F_2(-\bq_{12},\bk),
\nonumber\\
 T^{4h}(\bk,-\bk+\bq_{12},\bk',-\bk'-\bq_{12})&\simeq & 
 T^{4h}(\bk,-\bk,\bk',-\bk')
\nonumber\\
&&\hspace{-10em}+8 [ I^1_1(k) I^1_1(k')] ^2
P^{\lin}(\bq_{12})P^{\lin}(k)P^{\lin}(k')
\left[F_2(\bq_{12},-\bk)F_2(\bq_{12},\bk')
+F_2(\bq_{12},-\bk)F_2(\bq_{12},-\bk')
\right],
\label{eq:Thm}
\end{eqnarray}
\hidden{Terms with $F_2$ need to be expanded since there is a pole for the long wavelength 
mode
\begin{eqnarray}
 T^{4h}(\bk,-\bk+\bq_{12},\bk',-\bk'-\bq_{12})
&\approx&  T^{4h}(\bk,-\bk,\bk',-\bk')+
4 [ P^{\non}(k) F_2(\bq_{12},-\bk) + P^{\non}(|\bk-\bq_{12}|) F_2(\bq_{12},\bk-\bq_{12}) ]
\nonumber\\
&& \times [  P^{\non}(k') F_2(-\bq_{12},-\bk') + P^{\non}(|\bk'+\bq_{12}|) F_2(-\bq_{12},\bk'+\bq_{12}) ]
\label{eq:Thm}
\end{eqnarray}
where $P^\non(k) = [I_1^1(k)]^2 P^\lin(k) $  [even neglecting the $q$ correction, original formula seems to have some sign flips]. }
where we have used  approximations such as $|\bk+\bq_1|\simeq k$.

Inserting the halo model expressions for the matter trispectrum into
Eq.~(\ref{eq:pscov}), we  find the non-Gaussian term can be broken up as
\begin{equation}
C^{\rm NG}_{ij}=C_{ij}^{T0} + C^{\rm SSC} ,
\end{equation}
where $C_{ij}^{T0}$ was given in Eq.~(\ref{eq:pscov_T0}) and is the
standard non-Gaussian term involving only oppositely directed
Fourier modes \cite{Scoccimarroetal:99}.  
The last term $C_{ij}^{\rm SSC}$ is the
sum of 
the pieces involving $P^{\lin}(q_{12})$ in the 
$T^{2h}_{22}$, $T^{3h}$, and $T^{4h}$ terms of Eq.~(\ref{eq:Thm}).
We shall see that 
{the combined terms}
are exactly the super-sample covariance
term of Eq.~(\ref{eq:covcon}).

To see this relationship let us begin with the  remaining $P^\lin$ piece of the
 $T_{22}^{2h}$ term.   It can be simplified as 
\begin{eqnarray}
C^{\rm HSV}_{ij}&=&
\frac{1}{V_W^2}\int_{|\bk|\in k_i}\!\!\frac{d^3\bk}{V_{k_i}}
\int_{|\bk|'\in k_j}\!\!\frac{d^3\bk'}{V_{k_j}}
I^1_{2}(k,k)I^1_{2}(k',k')
\int\!\!\left[\prod_{a=1}^4\frac{d^3\bq_a}{(2\pi)^3}
\tW(\bq_a)
\right](2\pi)^3\delta^3_D(\bq_{1234})P^{\lin}(\bq_{12})
\nonumber\\
&=&\frac{1}{V_W^2}\int_{|\bk|\in k_i}\!\!\frac{d^3\bk}{V_{k_i}}
\int_{|\bk|'\in k_j}\!\!\frac{d^3\bk'}{V_{k_j}}
I^1_{2}(k,k)I^1_{2}(k',k')
\int\!\!\frac{d^3\bq_1}{(2\pi)^3}
\frac{d^3\bq_2}{(2\pi)^3}
\frac{d^3\bq}{(2\pi)^3}
\tW(\bq_1)\tW(\bq-\bq_1)\tW(\bq_2)\tW(-\bq-\bq_2)P^{\lin}(q)
\nonumber\\
&\simeq & 
I^1_{2}(k_i,k_i)I^1_{2}(k_j,k_j)\frac{1}{V_W^2}
\int\!\!\frac{d^3\bq}{(2\pi)^3}
|\tW(\bq)|^2P^{\lin}(q),
\label{eq:cov_hsv}
\end{eqnarray}
\hidden{note sign on $q$ in $W$'s,  the $d^2 \rightarrow d^3$, these issues persist in the eqns below too}
where we have used the variable change, 
$\bq_{12}\rightarrow \bq$, in the 2nd line on the r.h.s., and we have
use the identity (Eq.~\ref{eq:WWk}) for the window function.   Note that the remaining 
integral over $P^{\lin}$ is simply the variance of the linear density field convolved with the window
function,  Eq.~(\ref{eq:sigmw}).   The HSV term can now be simply expressed as
\begin{equation}
 {C}^{\rm HSV}_{ij}=(\sigma_W^L)^2
I^1_{2}(k_i,k_i)I^1_{2}(k_j,k_j). 
\end{equation}
We have labeled this term ``HSV'' for
halo sample variance as it takes the form originally pointed out in 
Ref.~\cite{Satoetal:09} in their Eq.~(18)
\citep[see also Section~3.4 in Ref.][]{Kayoetal:12},
although those works focused on the two-dimensional case (i.e.\ lensing
power spectrum).

The super-sample covariance consistency relation Eq.~(\ref{eq:covcon}) is
now manifest with Eq.~(\ref{eq:Iconsistency}) in the 1-halo dominated regime
\begin{equation}
\frac{\partial \ln P(k)}{\partial \delta_b} \approx \frac{I_2^1(k,k)}{I_2^0(k,k)}.
\label{eq:HSVresponse}
\end{equation}
In the halo model this consistency arises from compatibility of the bias with the mass function. 
The number density of rare halos responds to the background mode more strongly according
to its bias and so the response function Eq.~(\ref{eq:HSVresponse}) involves the average
bias of the halos contributing to the bin in $k$.

Next let us consider the remaining $P^\lin$ piece to the $T^{4h}$ term.   
This can be similarly simplified as 
\begin{eqnarray}
 C^{\rm BC}_{ij}&=&\frac{1}{V_W^2}
\int_{|\bk|\in k_i}\!\!\frac{d^2\bk}{V_{k_i}}
\int_{|\bk|'\in k_j}\!\!\frac{d^2\bk'}{V_{k_j}}
\int\!\!\frac{d^3\bq_1}{(2\pi)^3}
\frac{d^3\bq_2}{(2\pi)^3}
\frac{d^3\bq}{(2\pi)^3}\tW(\bq_1)\tW(\bq-\bq_1)\tW(\bq_2)\tW(\bq-\bq_2)P^{\lin}(q)
\nonumber\\
&&\hspace{10em}\times 
\left\{
8[ I^1_1(k) I^1_1(k')]^2P^{\lin}(k)P^{\lin}(k') 
\left[F_2(\bq,-\bk)F_2(\bq,\bk')
+F_2(\bq,-\bk)F_2(\bq,-\bk')
\right]\right\}\nonumber\\
&\simeq&\left(\frac{68}{21}\right)^2  [I^1_1(k_i)
I^1_1(k_j)]^2P^{\lin}(k_i)P^{\lin}(k_j) \, (\sigma_W^L)^2,
\label{eq:cov_bc}
\end{eqnarray}
where we have used the identity (Eq.~\ref{eq:WWk}) for the window
function and used the azimuthal-angle average of the perturbation
theory kernel $F_2$ as
\begin{equation}
 \int_{-1}^{1}\!\frac{d\mu_{12}}{2}F_2(\bq_1,\bq_2)
=\frac{5}{7}+\frac{2}{21}=\frac{17}{21}.
\end{equation}
\hidden{Note that the pole only vanishes on azimuthal averaging indicating that higher order
terms in a $q/k$ expansion are important.   Expanding the correction
\begin{equation}
{\frac{68}{21}} \rightarrow \frac{47}{21} - \frac{1}{3}\frac{d \ln P^{\non}}{d\ln k}
\end{equation}
compare this with Sherwin \& Zaldarriaga eq. 24 for squeezed bispectrum.  The power spectrum derivative term appears here since this term corresponds to a dilation of the volume and its impact depends on the slope of $P(k)$ or more transparently the dimensionless $k^3 P(k)$.
\begin{equation}
{\frac{68}{21}} \rightarrow \frac{68}{21} - \frac{1}{3}\frac{d \ln k^3 P^{\non}}{d\ln k}
\end{equation}
This is also what one would get from taking a separate universe calculation
for 
\begin{equation}
\frac{\partial \ln    P(k)}{\partial  \delta_b} 
\end{equation}
if the derivative is taken at fixed physical scale or Eulerian coordinates rather than fixed ``comoving'' or Lagrangian coordinates.    I think the halo terms already take that shift into account by absorbing the
transformation into Eulerian bias.  Mathematically, those terms don't have poles to worry about.}
This type of term was called beat coupling (BC) in Ref.~\cite{Hamiltonetal:06}.  Our expression differs from their
Eq.~(94) only through the inclusion of the $I_1^1$ terms, which mediate
the transition to the 1 halo regime.
Ref.~\cite{Hamiltonetal:06} also treated beat coupling in the deeply nonlinear regime by employing a model
of the nonlinear matter trispectrum based on 
hyper extended perturbation theory \cite{ScoccimarroFrieman:99} and
derived the similar formula to Eq.~(\ref{eq:cov_bc}) for the nonlinear
regime, where the covariance terms have the nonlinear matter power
spectra instead of the linear spectra and have a different prefactor
from 17/21. 
However,  the halo model differs from hyper extended
perturbation theory in the nonlinear regime, and 
Refs.~\cite{TakadaJain:09} and \cite{Satoetal:09} showed 
that the HSV effect based on
the halo model (Eq.~\ref{eq:cov_hsv}) gives a much better agreement to
the power spectrum covariance in the nonlinear regime measured from
the simulations of weak lensing field than predicted from 
the BC effect using the
hyper extended perturbation theory. For this reason, we will hereafter use
the label ``BC''  to refer the super-sample covariance in the weakly
nonlinear regime only.

With these simplifications, the BC term is  also exactly what we obtain from the consistency relation
in the perturbative regime where the 2 halo term dominates the power spectrum.  The 
background response,
\begin{equation}
\frac{\partial \ln    P(k)}{\partial  \delta_b} = \frac{68}{21},
\end{equation}
can be derived by considering the growth in a separate universe with rescaled mean density in the  Einstein-de Sitter approximation that matches the assumptions for the $F_2$ kernel
(see Eq.~122 in Ref.~\cite{Baldauf:2011bh}).
%

The final term is the $P^L$ piece of the $T^{3h}$ term:  
\begin{eqnarray}
 C^{\rm HSV-BC}_{ij}&=&\frac{1}{V_W^2}
\int_{|\bk|\in k_i}\!\!\frac{d^2\bk}{V_{k_i}}
\int_{|\bk|'\in k_j}\!\!\frac{d^2\bk'}{V_{k_j}}
\int\!\!\frac{d^3\bq_1}{(2\pi)^3}
\frac{d^3\bq_2}{(2\pi)^3}
\frac{d^3\bq}{(2\pi)^3}\tW(\bq_1)\tW(\bq-\bq_1)\tW(\bq_2)\tW(\bq-\bq_2)P^{\lin}(q)
\nonumber\\
&&\hspace{10em}\times 
\left\{
4[I_1^1(k)]^2I^2_1(k',k')P^{\lin}(k)F_2(\bq,\bk')
+4[I_1^1(k')]^2I^2_1(k,k)P^{\lin}(k')F_2(\bq,\bk)\right\}\nonumber\\
&\simeq& 
(\sigma_W^L)^2
\left\{
\frac{68}{21}[I^{1}_1(k_i)]^2I^1_2(k_j,k_j)P^{\lin}(k_i)
+ ( {i \leftrightarrow j})
\right\}.
\end{eqnarray}
This contribution takes the form of the cross term between HSV and BC.

We can
now see that the sum of all three terms 
\begin{equation}
C_{ij}^{\rm SSC} = C_{ij}^{\rm HSV} + C_{ij}^{\rm  HSV-BC} + C_{ij}^{\rm
 BC}
\label{eq:cov_ssc}
\end{equation}
reproduces the general expression for the super-sample covariance of Eq.~(\ref{eq:covcon})
with the total halo model response
\begin{equation}
\frac{\partial \ln  P(k)}{\partial  \delta_b} \approx \frac{\frac{68}{21}  [I_1^1(k)]^2 P^\lin(k) + I_2^1(k,k)}{[I_1^1(k)]^2 P^\lin(k) + I_2^0(k,k)} .
\label{eq:haloresponse}
\end{equation}
Thus the halo model  obeys the consistency relation for the trispectrum
of Eq.~(\ref{eq:consistency}) and illustrates that super-sample covariance can be described
by the response of the power spectrum to the background density.   Beyond the halo model,
it can be accurately calibrated directly from simulations.

\begin{figure}
\begin{center}
\includegraphics[height=8.7cm,angle=-90]{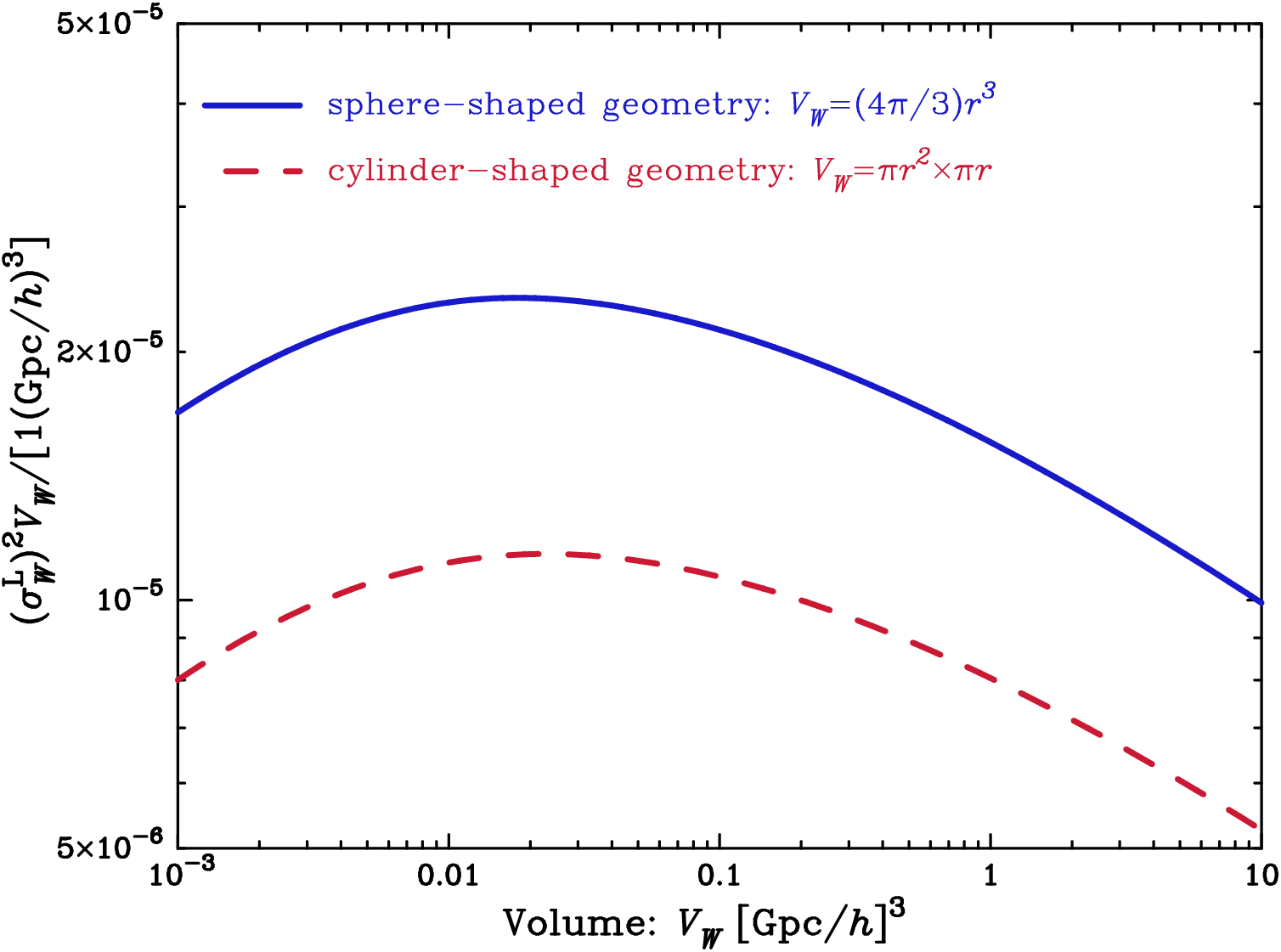}
\includegraphics[height=8.7cm,angle=-90]{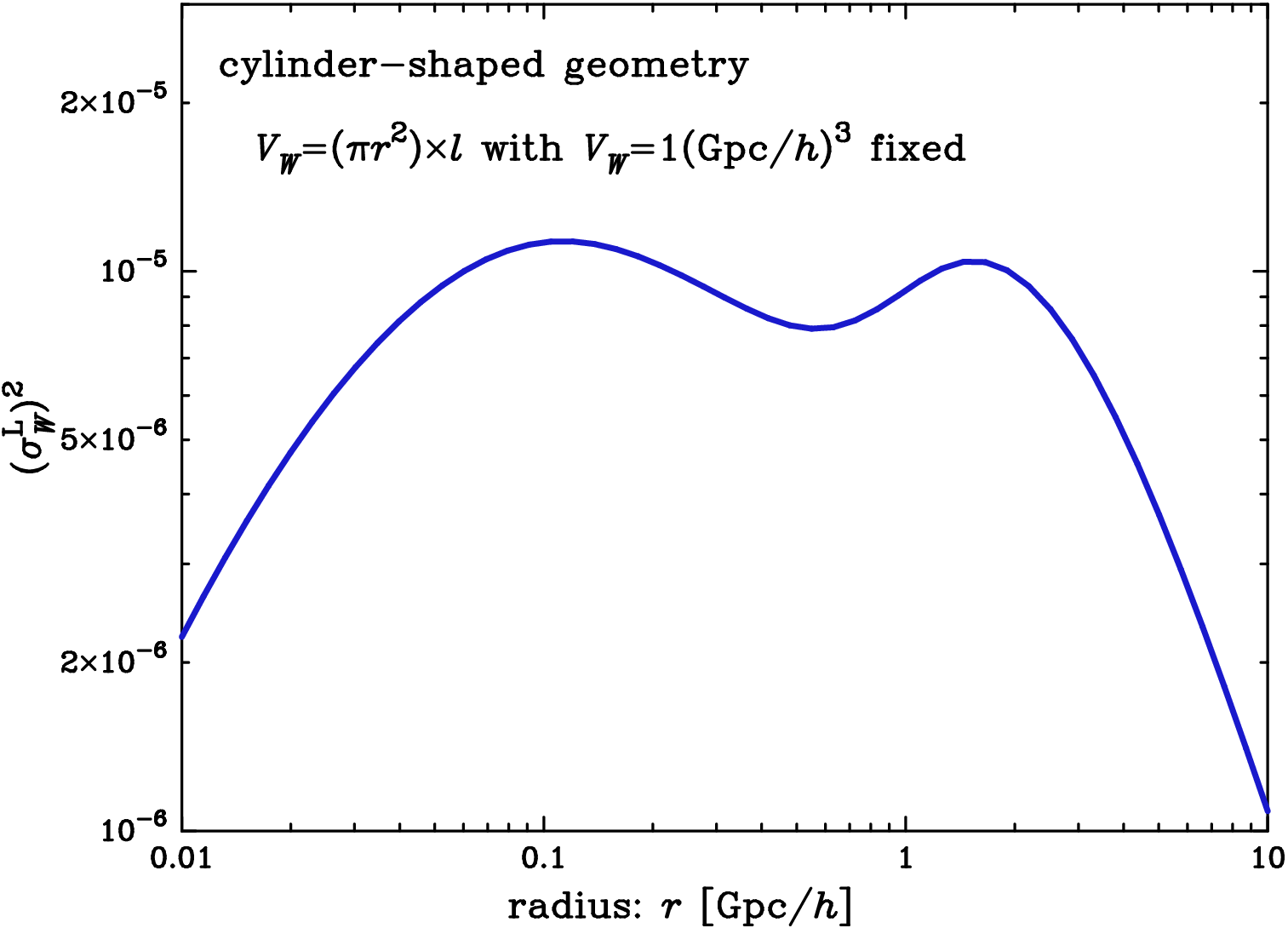}
\caption{\label{fig:sigw}  The variance of the linear mass density field
 convolved with the survey window function, $(\sigma_W^L)^2$  for a $\Lambda$CDM model and $z=0$.   {\it Left}: variance
 as a function of survey volume  for spherical $V_W=(4\pi/3)r^3$
 and cylindrical windows $V_W=\pi r^2 l; l=\pi r$.  We show the variance multiplied by the volume  to illustrate that for
 a wide range in survey volumes the relative impact of super-sample variance
 and the other terms which scale as $1/V_W$ remains the same within a factor of 2 due to the flatness of
 $P^\lin(k)$ around matter-radiation equality.
  {\it Right}:  cylindrical windows of fixed volume $V_W=\pi r^2 l = 1 ($Gpc$/h)^3$.   Optimizing for
 both minimum variance and a non-elongated geometry (see text) yields $l=\pi r$ as the best choice for
 volumes that exceed the matter-radiation turnover.   As shown in the left panel, this choice has a lower variance
 than the spherical window of the same volume.}
\end{center}
\end{figure}

\begin{figure}
\begin{center}
\includegraphics[height=8.7cm,angle=-90]{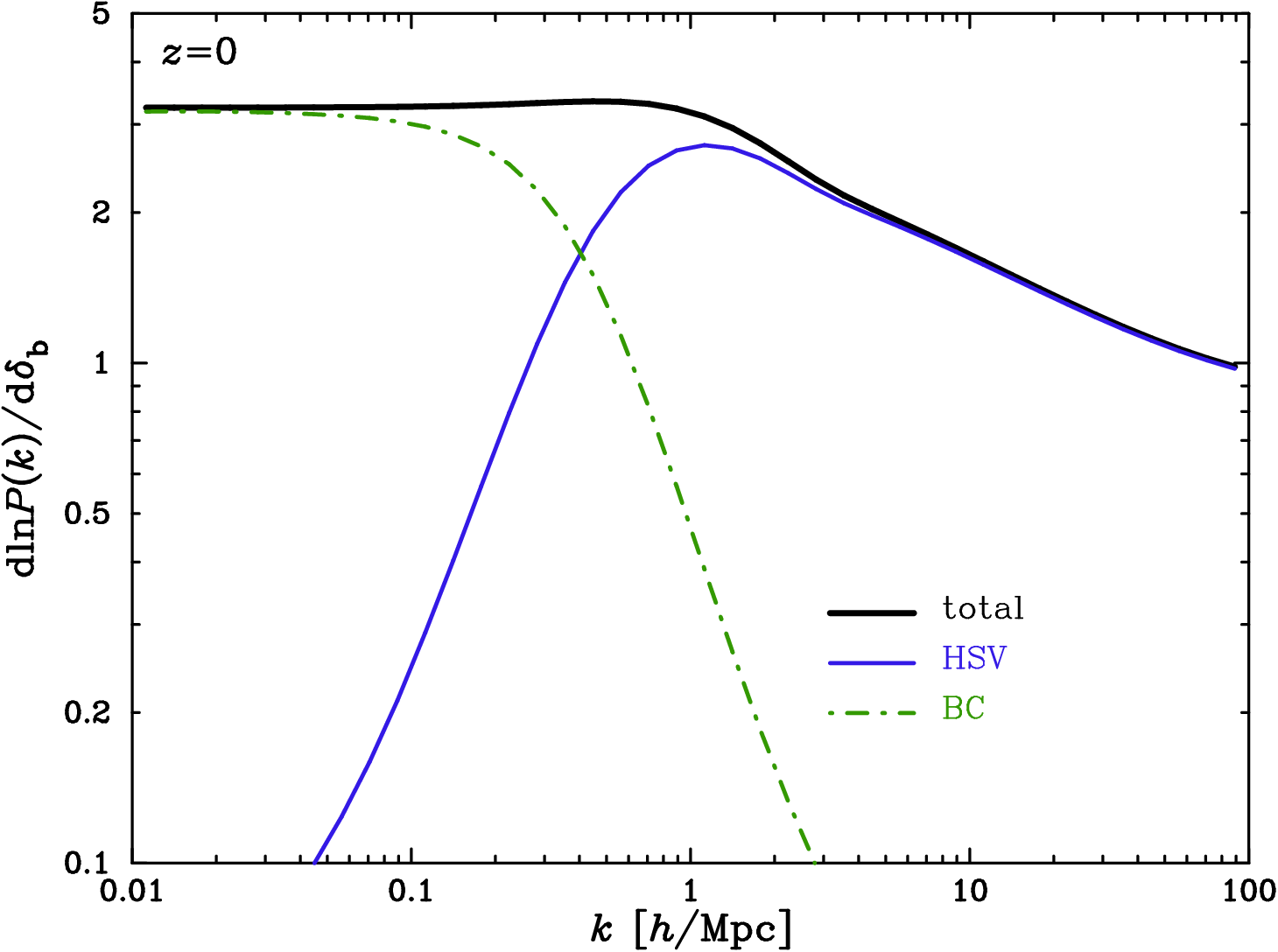}
\includegraphics[height=8.7cm,angle=-90]{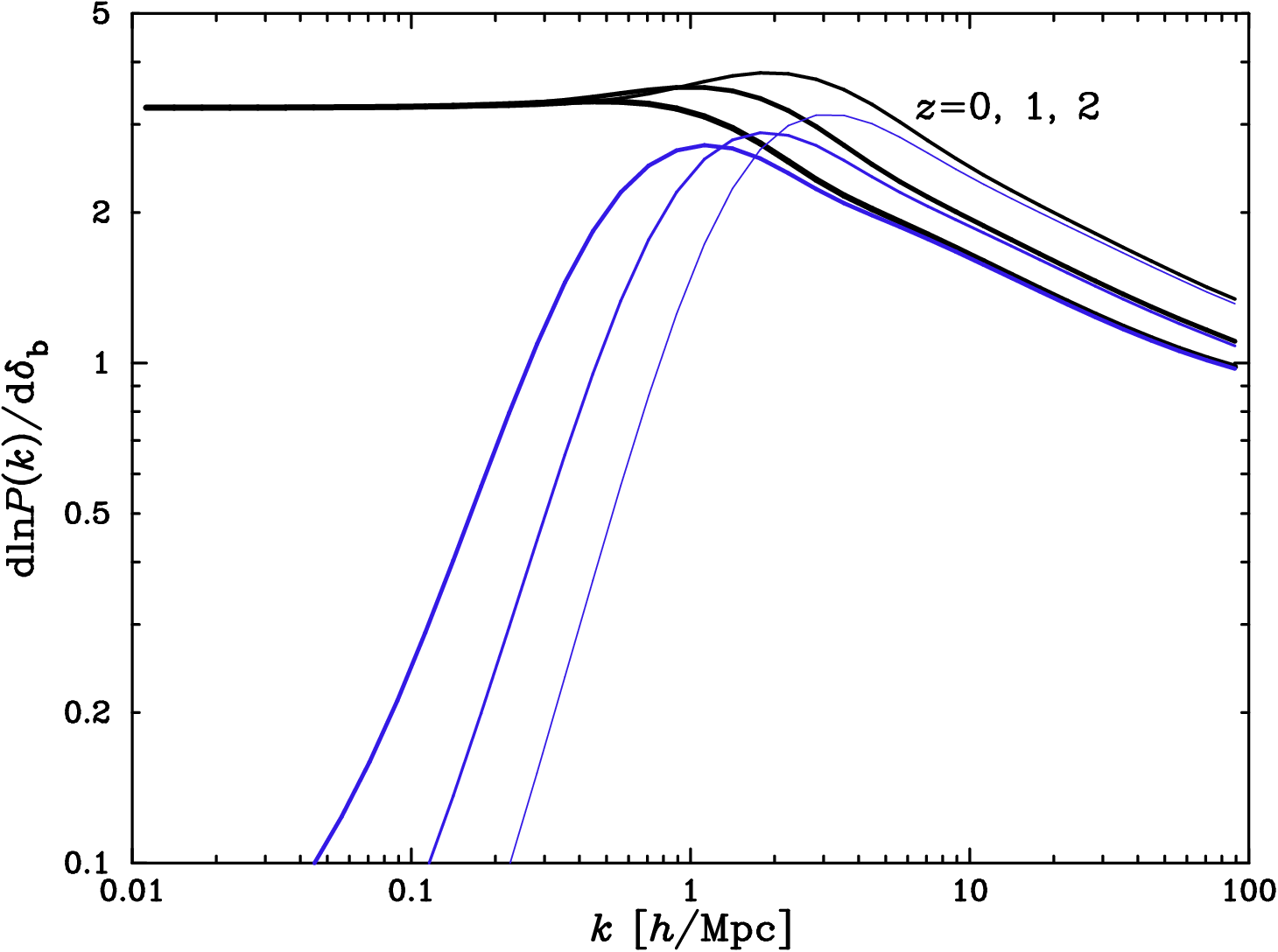}
\caption{
The response function of the power spectrum to the super-survey
mode, $\partial \ln P(k)/\partial \delta_b$ for the halo model
 (Eq.~\ref{eq:haloresponse}). {\it Left:} $z=0$.
Thin solid and dot-dashed curves are the halo sample variance
 and the beat-coupling term respectively. The total power
 shows a plateau up to $k\sim 1~h/{\rm Mpc}$, having $\partial \ln
 P/\partial \delta_b\simeq 3$ for the amplitude.  
{\it Right:} $z=0, 1$ and $2$.  At higher redshifts the
 plateau develops a small ridge due to the  larger response of more rare halos.
 \label{fig:dlnPddb}
}
\end{center}
\end{figure}

\subsection{$\Lambda$CDM examples}
\label{sec:LCDM}

To illustrate the power spectrum covariance for a $\Lambda$CDM
model, we employ cosmological parameters that are consistent with the
WMAP 9-year result \cite{WMAP9}: $\Omega_{\rm m0}h^2=0.136$,
$\Omega_{\rm b0}h^2=0.0226$ and $\Omega_{\rm m0}=0.278$ for the density
parameters assuming a flat universe and $n_s=0.972$ and $A_s=2.41\times
10^{-9}$ at $k=0.002~{\rm Mpc}^{-1}$ for the primordial power spectrum parameters. We use the
fitting formula in Ref.~\cite{EisensteinHu:99} to compute the transfer
function for the model. Note $\sigma_8=0.83$. As for the halo model
ingredients, the halo mass function, the halo mass profile, and the halo
bias, we used the same modes as in Ref.~\cite{OguriTakada:10}.

The fundamental building block for the super-sample covariance effect is the
variance of the density field averaged over the survey window,
$(\sigma_W^L)^2$.  Our  approach is not limited to simple
window geometries since the power spectrum at high $k$ should respond to
the background density in the same way regardless of geometry. Here we
consider spherical- or cylinder-shaped geometries as working examples for which
the window functions are
\begin{align}
 |\tW(\bk)|= & 3\frac{j_1(k r)}{k r} V_W ,  \qquad  V_W=(4\pi/3)r^3, \nonumber\\
   |\tW(\bk)| =& 2\frac{J_1(k_\perp
r)}{k_\perp r}\frac{\sin(k_\parallel l/2)}{k_\parallel l/2} V_W, \qquad V_W=(\pi r^2) l,
\end{align}
where 
$k^2=k_\perp^2+k_\parallel^2$.
The left panel of
 Fig.~\ref{fig:sigw} 
shows how the variance
scales with survey window volume compared to $1/V_W$, which is the scaling of
the standard covariance terms in T0  {(Eqs.~\ref{eq:pscov_org}
and \ref{eq:pscov_T0}).}
The main result is
that for both geometries {the scaling of $(\sigma_W^L)^2$}
differs from the standard scaling
only by a factor of 2 across the whole 4 orders of 
magnitude 
%
in volume.
This is because the matter power spectrum $P^\lin(k)$ is relatively flat
on the corresponding scales.   It is only for volumes significantly greater than $1$ (Gpc/$h)^3$ that
the super-sample variance effect declines relative to the other terms. 

For a cylinder window, we employ a specific shape given by $l=\pi
r$.  This is motivated by the study in the right panel of Fig.~\ref{fig:sigw} for the impact of the
aspect ratio $r/l$ or $r$ at a fixed volume $1$ (Gpc/$h)^3$.  
For a general aspect ratio and volume, the variance scales as $k_\parallel k_\perp^2 P^{\lin}(k)$.   Since the maximum $k_\parallel k_\perp^2 \propto l^{-1} r^{-2} =$ const., minimizing the
variance involves minimizing $P^{\lin}(k)$ at the maximum $k$.   Increasing from small $r$ or a ``tube'' geometry, $k \propto 1/r$
and so the variance increases until the peak of the power spectrum is reached.  It then declines
until the minimum $k$ is achieved when $l \approx \pi r$.   The volume then becomes flattened into
a ``pill'' geometry where $k \sim 1/l \propto r^{-2}$.   The variance increases until the peak of $P^{\lin}(k)$ is
crossed in the opposite direction and then declines thereafter.  Since either the extreme ``tube'' or ``pill'' cases have undesirable
effects of washing out structure in $P^{\lin}(k)$ via the convolution with the window and the smallest 
dimension cannot be in the nonlinear regime, this implies that for a cylindrical geometry the best
shape for minimizing super-sample variance is $l \approx \pi r$.   At this minimum point, the
cylinder has a smaller variance by a factor of $\sim 2$ compared with a
sphere of the same volume.  

The second building block of the super-sample covariance effect is the response of the nonlinear
$P(k)$ to a fractional change in the background density $\delta_b$.   
In Fig.~\ref{fig:dlnPddb}, we study this response function $\partial \ln P/\partial
\delta_b$ as predicted by the halo model {(Eq.~\ref{eq:haloresponse}).}  
In the one and two halo regimes, the response function is given by the HSV and BC effects respectively.   Summing up the two terms gives an plateau-like feature in
the response function up to a certain wavenumber, $k\sim 1~h/{\rm Mpc}$
for $z=0$ or somewhat smaller wavenumbers for the higher redshift. This
implies that the super-sample variance  can be
absorbed by a multiplicative, constant factor in the power spectrum
amplitude, up to the certain wavenumber.  In the halo model, 
higher wavenumber corresponds to
smaller, less biased halos where the response decreases.   Since halos of a given size are more
rare at high redshift, there is an increase in response at higher redshift which creates a ridge at the transition from
BC to HSV domination.

\begin{figure}
\begin{center}
\includegraphics[width=10.cm,angle=-90]{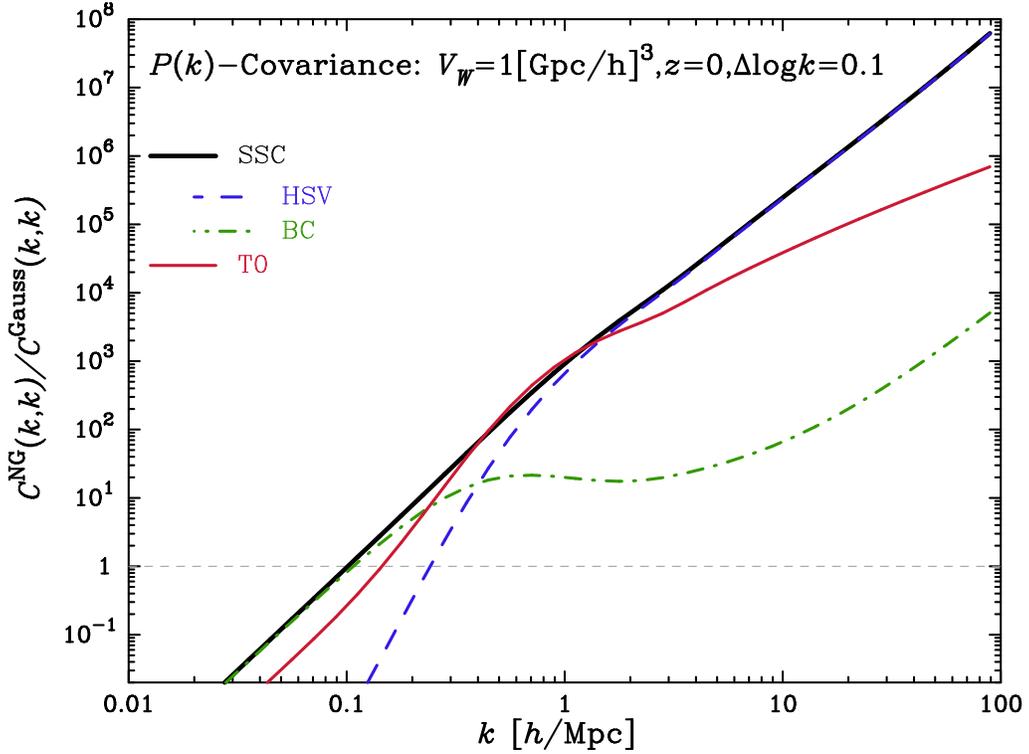}
\caption{The diagonal term of 
power spectrum covariance as a function of $k$
in a $\Lambda$CDM model for a spherical survey volume of
$V_W=1~({\rm Gpc}/h)^3$,  redshift $z=0$, and  $k$-binning of 
$\Delta\ln k=0.1$,  respectively. Plotted here is the non-Gaussian
 covariance contributions relative to the Gaussian term. At $k\simgt 0.1~h/{\rm Mpc}$,
 the non-Gaussian errors start to dominate over the Gaussian term for
 the $k$-binning. The bold solid curve is the halo model prediction for
the super-sample covariance (SSC), which is broken down into the dominant components
on the weakly nonlinear and fully nonlinear scales respectively: 
the halo sample variance (HSV; dashed) and
the beat-coupling effect (BC; dot-dashed)
(see Eq.~\ref{eq:cov_ssc}).
The red (light) solid  curve is the standard trispectrum term (T0) which is subdominant
or comparable throughout.}
 \label{fig:pkcov}
\end{center}
\end{figure}

We can now put these two pieces together to form the contribution to the power spectrum covariance
from the super-sample covariance terms.
Fig.~\ref{fig:pkcov} shows the non-Gaussian diagonal, or variance, term relative to the Gaussian expectation for a $V_W=1$ (Gpc/$h)^3$ spherical survey with
bins of $\Delta \log k=0.1$.  As expected, the non-Gaussian contributions are larger than Gaussian for $k\simgt 0.1~h/{\rm Mpc}$ 
where the linear to nonlinear transition occurs or equivalently where the 2-halo to 1-halo transition occurs.   
Where the curve crosses unity 
specifically depends on the binning scheme since the Gaussian variance terms are suppressed as the
number of independent $k$-modes increases.

For the halo model, the BC term dominates over other non-Gaussian contributions 
in the {weakly nonlinear}
regime and the HSV term in the {deeply} 
nonlinear regime with a smooth transition in between, {which is} 
mediated by the cross term.  The total SSC contribution 
therefore dominates or is comparable to the standard T0 term  everywhere.

These results are of course limited by the accuracy of the halo model
itself 
which does not 
directly suffice for future surveys.
  Weak lensing
cosmology, which is the primary science driver of planned imaging
surveys, needs to use the power spectrum information up to $k\sim
1~h/{\rm Mpc}$ to attain the full potential
\cite{HutererTakada:05}. 
Galaxy clustering based cosmology aims at using
the information up to at least  a few $0.1$~Mpc$/h$ to capture the
baryon acoustic oscillations as well as to measure the redshift-space
distortion effect \cite{Takadaetal:06,Ellisetal:12}. 
Thus an accurate  simulation-based calibration of the
background response $d \ln P/d\delta_b$ across this regime is important and will be presented in a
separate paper.

\section{Discussion}
\label{sec:discussion}

In this paper we have developed a simple, unified approach to describe 
the super-sample covariance of power spectrum estimation
in a finite-volume survey.
We show that the previously known effects of ``beat coupling" and ``halo sample variance"
are both just limiting cases of the general response of the power spectrum to a change
in the background density.

Formally, all power spectrum covariance effects are described by the matter trispectrum due to the two-point
nature of power spectrum estimators.   
The super-sample covariance term arises due to the convolution of modes across the Fourier
width of the survey window or ``squeezed'' trispectrum configurations which connect
pairs of closely separated short-wavelength modes through a long-wavelength or super-survey
mode.   We show that there exists a consistency relation between these squeezed trispectra
and the response of the power spectrum to a change in the background density.
This consistency relation also exposes
why this term is a {\it co}variance: the power spectrum responds coherently across bins
to a single unknown background density.   Our formulation is general and applicable to any survey geometry
since the response to a background mode does not depend on the detailed
geometry of the window. 

To make contact with the literature, we used the halo model trispectrum to illustrate
these effects.     We find that in the weakly nonlinear regime the response is exactly what is
known as beat coupling.  In the fully-nonlinear regime it is exactly what is known
as halo sample variance.   The joint effect is the dominant non-Gaussian covariance
for a wide range of survey volumes and its accurate calibration will be important
for  
high-precision cosmology using large-scale structure probes. 
Our description also exposes the fact that accurate calibration is straightforward.   
Since the effect is the response of the power spectrum to a change in the background density,
it only requires running two simulations with different background parameters to calibrate at any given redshift.

Our construction also exposes the possibility that power spectrum super-sample covariance need
not be considered as an additional source of noise at all, but rather an additional signal of known shape but
unknown amplitude given by the background density in the survey.   By including it
in the covariance, one pre-marginalizes the impact of the unknown amplitude.   
Alternately, one can fit for its amplitude given the template power spectrum response
under a prior given by its expected variance in the survey window.  
This approach also directly exposes its impact on cosmological parameter estimation
as different parameters will have different degeneracies with this background mode
(see Ref.~\cite{TakadaJain:09} for the similar
 discussion).
  It can also
protect against unlikely realizations of the data by examining the dependence of 
results on the prior.  

In principle, a joint fit could recover some information on super-survey modes from
the observable sub-survey power spectrum albeit limited by degeneracies with 
global cosmological parameters.  
 This opens up an interesting possibility to explore
very large-wavelength fluctuations based on the background mode-coupling to short-wavelength
observables.

\begin{acknowledgements}
We thank Gary Bernstein, Nick Gnedin, Bhuvnesh Jain, Issha Kayo, 
Elisabeth Krause,
 Andrey Kravtsov, Marilena Loverde, Surhud More, Roland de Putter, 
Ryuiichi Takahashi 
and Jochen Weller for useful discussions.  
MT was supported by World
 Premier International Research Center Initiative (WPI Initiative),
 MEXT, Japan, by the FIRST program ``Subaru Measurements of Images and
 Redshifts (SuMIRe)'', CSTP, Japan, and by Grant-in-Aid for
 Scientific Research from the JSPS Promotion of Science (23340061).
WH was
 supported by U.S.\ Dept.\ of Energy contract DE-FG02-90ER-40560, the
 David and Lucile Packard Foundation, and the Kavli Institute for
 Cosmological Physics at the University of Chicago through grants NSF
 PHY-0114422 and NSF PHY-0551142 and an endowment from the Kavli
 Foundation and its founder Fred Kavli.  WH acknowledges the hospitality of
 Kavli IPMU where this work was initiated.  
\end{acknowledgements}

{\em Note Added:}  Following publication, we discovered a  term in  the linear regime
response to a background mode missing in Ref.~\cite{Hamiltonetal:06} and subsequent treatments (see arXiv:1401.0385).   Eq.~(\ref{eq:haloresponse})
should be replaced with
\begin{equation}
\frac{\partial \ln  P(k)}{\partial  \delta_b} \approx \frac{
\left( \frac{68}{21} -\frac{1}{3} d\ln k^3  [I_1^1(k)]^2 P^\lin(k)/d\ln k \right)  [I_1^1(k)]^2 P^\lin(k)  + I_2^1(k,k)}{[I_1^1(k)]^2 P^\lin(k) + I_2^0(k,k)} ,
\label{eq:haloresponsecorrected}
\end{equation}
which lowers the response in the linear regime and leaves it unchanged in the nonlinear regime.
\appendix

\section{Power spectrum covariance for projected density fields}
\label{app:2d}

The covariance formalism developed in the main paper can be directly applied
to any statistic that is derived from the matter power spectrum.   In particular, many applications
such as weak lensing involve a weighted 2D projection of the density field.   We use the flat-sky approach
developed in Refs.~\cite{CoorayHu:01,TakadaJain:09} for the two point statistics of a single projected field
but the formalism can easily be generalized to cross spectra of multiple fields 
\cite{Hu:2003pt,Baldaufetal:10,OguriTakada:10} or all-sky statistics \cite{Hu:2001fb,Sehgaletal:10,dePutterTakada:10,Becker:12}.

We define the projected density
field as
\begin{equation}
 \Sigma(\btheta)=\int\!d\chi~f(\chi)\delta(\chi\btheta,\chi),
\end{equation}
where we have assumed a spatially flat  universe with radial coordinate
$\chi(z)=\int_0^z dz'/H(z)$ and $f(\chi)$ is the
radial selection function. For the weak lensing field, $f(\chi)$ is
given by Eq.~(4) in Ref.~\cite{TakadaJain:09}. 
Taking into account the survey window function, the observed field is
given as
\begin{equation}
 \Sigma_\wt(\btheta)=\wt(\btheta)\Sigma(\btheta),
\end{equation}
where $\wt(\btheta)$ is the survey window function on the sky;
$\wt(\btheta)=1$ if the pixel in the direction $\btheta$ on the sky is
in the survey region or contains data, otherwise $\wt(\btheta)=0$. 

Under the flat-sky approximation, the Fourier-transformed field
becomes
\begin{equation}
 \tsigma_\wt(\bl)=\int\!\frac{d^2\bl'}{(2\pi)^2}\twt(\bl-\bl')
\tsigma(\bl').
\end{equation}
Similarly to Eq.~(\ref{eq:ps_est}),
the power spectrum estimator for the projected field is defined as
\begin{equation}
 \hat{C}(l_i)\equiv \frac{1}{\Omega_\wt}\int_{|\bl|\in
  l_i}\!\frac{d^2\bl}{\Omega_{l_i}}\tsigma_\wt(\bl)\tsigma_\wt(-\bl),
\end{equation}
where $\Omega_\wt$ is the effective survey area defined as $\Omega_\wt\equiv
\int\!d^2\btheta~ \wt(\btheta)$, and $\Omega_{l_i}=\int_{|\bl|\in
l_i}\!d^2\bl \simeq 2\pi l_i\Delta l$ when $\Delta l / l_i\ll 1$.

In the limit that the angular mode of $l$ is much greater than the width
of the window function, the power spectrum estimator is unbiased in a
sense that the ensemble average gives the underlying true power
spectrum:
\begin{equation}
 \ave{\hat{C}(l_i)}=C(l_i). 
\end{equation}
%
As in the 3D case, we can derive the covariance of 
the 2D power spectrum including the effect of the window
function:
%
\begin{eqnarray}
 \Ca_{ij} &\equiv&
  \ave{\hat{C}(l_i)\hat{C}(l_j)}-C(l_i)C(l_j)\nonumber\\
&=&\frac{1}{\Omega_\wt}\left[
\frac{(2\pi)^2}{\Omega_{l_i}}C(l_i)\delta^K_{ij}+ \bar \Ta^\wt(l_i,l_j) \right],
\end{eqnarray}
where
\begin{eqnarray}
\bar \Ta^\wt(l_i,l_j) = \frac{1}{\Omega_\wt}  
\int_{|\bl|\in l_i}\!\frac{d^2\bl}{\Omega_{l_i}}
\int_{|\bl'|\in l_j}\!\frac{d^2\bl'}{\Omega_{l_j}}
\int\!\left[\prod_{a=1}^4\frac{d^2\bq_a}{(2\pi)^2}
\tW(\bq_a)\right](2\pi)^2\delta^2_D(\bq_{1234})
\Ta(\bl+\bq_1,-\bl+\bq_2,\bl'+\bq_3,-\bl'+\bq_4).
\label{eq:pscov2D}
\end{eqnarray}
Using the Limber approximation, the angular power spectrum and trispectrum can be related to their 3D counterparts via
\begin{eqnarray}
C(l) &\approx & \int d\chi f^2(\chi) \chi^{-2} P(\bk;\chi), \nonumber\\
 \Ta(\bl_1,\bl_2,\bl_3,\bl_4) &\approx  &
\int\!d\chi f^4(\chi) \chi^{-6} T(\bk_1,\bk_2,\bk_3,\bk_4;\chi),
\end{eqnarray}
where $\bk_i=\bl_i/\chi$.

The trispectrum consistency relation then implies
\begin{equation}
\Ca_{ij} = \Ca_{ij}^{\rm G}+ \Ca_{ij}^{T0} + \Ca_{ij}^{\rm SSC}, 
\end{equation}
where
\begin{eqnarray}
\Ca_{ij}^{\rm G} &=&\frac{1}{\Omega_\wt}
\frac{(2\pi)^2}{\Omega_{l_i}}C(l_i)\delta^K_{ij}, \nonumber\\
\Ca_{ij}^{\rm T0} &=& \frac{1}{\Omega_\wt}
\int_{|\bl|\in l_i}\!\frac{d^2\bl}{\Omega_{l_i}}
\int_{|\bl'|\in l_j}\!\frac{d^2\bl'}{\Omega_{l_j}}
\Ta(\bl,-\bl,\bl',-\bl'), \nonumber\\
\Ca_{ij}^{\rm SSC} &=& \frac{1}{\Omega_\wt^2}
\int\!d\chi~
 f(\chi)^4 \chi^{-6}
\frac{\partial P(k_i;\chi) }{\partial \delta_b}
\frac{\partial P(k_j;\chi) }{\partial \delta_b}
\int\!\frac{d^2\bl}{(2\pi)^2}P^
\lin(k;\chi)|\twt(\bl)|^2,
\label{eq:pscov_2d}
\end{eqnarray}
where we have again used the Limber relation
$\bk_i=\bl_i/\chi$. 
With the halo model expressions for the response function from Eq.~(\ref{eq:haloresponse}) these
relations reproduce the results
in the previous work, Eq.~(18) in Ref.~\cite{Satoetal:09} and Eq.~(17)
in Ref.~\cite{TakadaJain:09} for the HSV and BC contributions, respectively.
  In principle, background modes also change the $k \leftrightarrow l$ relation through perturbations
to the distance redshift relation.   However such modes would have to be constant out to the
maximum distance in the integral and are typically much smaller than the
$\delta_b$ considered here.

\bibliography{refs}

\end{document}